\documentclass[twocolumn,prd,groupedaddress,nofootinbib,showpacs]
{revtex4}

\usepackage{latexsym}
\usepackage{amssymb}
\usepackage{graphicx}
\usepackage{dcolumn}
\usepackage{bm}

\begin{document}

\title{ Remarks on gauge fixing and BRST quantization of noncommutative 
gauge theories  }
\author{Ricardo Amorim$^{a}$, Henrique Boschi-Filho$^{b}$
and Nelson R. F. Braga$^{c}$ }
\affiliation{Instituto de F\'{\i}sica, 
Universidade Federal do Rio de Janeiro,
Caixa Postal 68528, RJ 21941-972 -- Brazil}

\date{\today}

\begin{abstract} 
We consider the BRST gauge fixing procedure of the noncommutative 
Yang-Mills theory and of the gauged $U(N)$ Proca model.
An extended Seiberg-Witten map involving ghosts, antighosts 
and auxiliary fields for non-Abelian gauge theories is studied.
We find that the extended map behaves differently for these models.
For the Yang-Mills theory in the Lorentz gauge it was not possible to find a map
that relates the gauge conditions in the noncommutative and ordinary 
theories. For the gauged Proca model we found a 
particular map relating the unitary gauge fixings in both formulations.
\end{abstract}

\maketitle

\section{Introduction}

\renewcommand{\theequation}{1.\arabic{equation}}
\setcounter{equation}{0}

Noncommutative gauge fields $Y$ can be described in terms of  ordinary 
gauge fields $y$ by using the Seiberg-Witten (SW) map\cite{SW} defined by 

\begin{equation}
\delta Y=\bar\delta Y[y]
\label{SW1}
\end{equation}

\noindent where $\delta $ and $\bar \delta$ are the gauge transformations 
for noncommutative and ordinary theories respectively and $ Y [y] $ means 
that we express the noncommutative fields $Y$ in terms of ordinary ones $y$. 
In this work we will consider gauge theories whose algebra closes as

\begin{equation}
[\delta_1,\delta_2]\,Y=\delta_3\, Y
\label{SW2a}
\end{equation}

\noindent without the use of equations of motion. 
In terms of the mapped quantities this condition reads
 
\begin{equation}
[\bar\delta_1,\bar\delta_2]\,Y[y]=\bar\delta_3\, Y[y]\,\,.
\label{SW2b}
\end{equation}

\noindent This implies that the mapped noncommutative gauge parameters 
must satisfy a composition law in such a way that they depend in general 
on the ordinary parameters and also on the fields $y$. 
Usually this noncommutative parameter composition law is the starting 
point for the construction of SW maps.
Interesting properties of Yang-Mills noncommutative theories where discussed 
in\cite{AK,Wess,REVIEW}, where these points are considered in detail. Other 
noncommutative theories with different gauge structures are also studied  in 
\cite{ABF,AFG}.

Originally the Seiberg-Witten map has been introduced relating 
noncommutative and ordinary gauge fields and the corresponding actions. 
When one considers the gauge fixing procedure
one enlarges the space of fields by introducing ghosts, antighosts and 
auxiliary fields. 
In this case one can define an enlarged BRST-SW map\cite{AF2} 

\begin{equation}
s\, Y\, =\, {\bar s}\, Y [y]
\label{0.0} 
\end{equation}

\noindent 
where $s$ and ${\bar s }$ are the BRST differentials for the noncommutative 
and ordinary theories respectively and $Y$ and $y$ here 
include ghosts, antighosts and 
auxiliary fields.  It is interesting to 
note that in this BRST approach the closure relation (\ref{SW2b}) is 
naturally contained in the above condition for the ghost field. 
This means that it is not necessary to construct a SW map for the parameter.
 
Here we will investigate the extension of this map also to the gauge fixed actions.
Observe that in ref.\cite{AF2} the Hamiltonian formalism was used while here we
consider a Lagrangian approach. 
In particular we study the consistency of such maps with the gauge fixing 
process. Considering the BRST quantization of a noncommutative theory we 
will find that some usual gauge choices for the  noncommutative theories 
are mapped in a non trivial way  in the  ordinary model.
 \medskip 

Recent results coming from string theory are motivating an increasing 
interest in noncommutative theories.  
The presence of an antisymmetric tensor background along the D-brane 
\cite{Po} world volumes (space time region where the string endpoints 
are located) is an important source for noncommutativity  in string 
theory\cite{HV,CH1}. Actually the idea that spacetime may be 
noncommutative at very small length scales is not new\cite{Heisenberg}. 
Originally this has been thought just as a mechanism for providing 
space with a natural cut off that would control ultraviolet 
divergences\cite{Snyder}, although these motivations have been eclipsed 
by the success of the renormalization procedures.

Gauge theories can be defined in noncommutative spaces by considering actions 
that are invariant under gauge transformations defined in terms of the Moyal 
structure\cite{SW}. 
In this case the form of the gauge transformations imply that the algebra of the 
generators must close under commutation and anticommutation relations. 
That is why  $U(N)$ is usually chosen as the symmetry group for
noncommutative extensions of Yang-Mills theories in place of $SU(N)$, 
although other symmetry structures can also be 
considered \cite{Wess}\cite{Bonora}\cite{AF}. 
Once one has constructed a noncommutative gauge theory,  it is possible to find the 
Seiberg-Witten map
relating the noncommutative fields to ordinary ones\cite{AK}.
The mapped  Lagrangian is usually  written as a nonlocal infinite series of 
ordinary fields and their space-time derivatives but the noncommutative 
Noether identities are however kept by the Seiberg-Witten map. This assures 
that the mapped theory is still gauge invariant.
 \medskip

In this work we will first consider (section {\bf II})
the case of the noncommutative $U(N)$ Yang-Mills 
theory and investigate the BRST gauge fixing procedure in the Lorentz gauge. 
Then the construction of the SW map between the noncommutative  and
ordinary Yang-Mills fields, including ghosts, antighosts and auxiliary fields
is discussed. 
We will see that imposing the Lorentz gauge in the noncommutative theory
does not imply an equivalent condition in the ordinary theory.
Conversely, imposing the Lorentz gauge condition in the 
ordinary gauge fields would not correspond to the same condition 
in the noncommutative theory. 

Another model that will present an interesting behavior is the gauged version 
of noncommutative non-Abelian Proca field, discussed in sections {\bf III}
and {\bf IV}.
For this model we study the BRST gauge fixing in the unitary gauge.
The model is obtained by introducing an auxiliary field which 
promotes the massive vector field to a gauge field.  
This auxiliary field can be seen as a pure gauge 
``compensating vector field" ,
defined in terms of $U(N)$ group elements and having a null curvature
\cite{CF,ABH}. In this model we find that the general SW  map  
relates in a non trivial way noncommutative and  ordinary gauge fixing conditions. 
However it is possible to find a particular SW map that relates 
the unitary gauges in noncommutative and ordinary theories. Some of these points 
have been partially considered in \cite{ABF}.

Regardless of these considerations, the Fradkin-Vilkovisky theorem\cite{FV} 
assures that the physics described by any non anomalous  gauge theory is 
independent of the gauge fixing, without the necessity of having the gauge fixing 
functions mapped. 
This means that the quantization can be implemented consistently  
in noncommutative and ordinary theories. 
Note that the SW map is defined for the gauge invariant action.
This places a relation between the noncommutative and ordinary theories
before any gauge fixing. 
Once a gauge fixing is chosen one does not  necessarily 
expect that the theories would still be related by the same map.
In the Yang-Mills case with Lorentz gauge we could not relate
the complete theories by a BRST-SW map after gauge fixing.
However, for the gauged Proca theory we could  find a map 
relating the unitary gauges in both noncommutative and ordinary theories.

\section{Gauge fixing the noncommutative $U(N)$ Yang-Mills theory}
\renewcommand{\theequation}{2.\arabic{equation}}
\setcounter{equation}{0}

In order to establish notations and conventions that will be useful later, 
let us start by considering the ordinary U(N) Yang-Mills action 
(denoting ordinary actions by the upper index $^{(0)}$)

\begin{equation}
S_0^{(0)}=\,tr \int d^{4}x\,\left(\,-\frac{1}{2}f_{\mu \nu } f^{\mu \nu }  \right)
 \label{1.1}
\end{equation}

\noindent 

where 

\begin{equation}
 f_{\mu\nu}=\partial_\mu a_\nu-\partial_\nu a_\mu-i[a_\mu,a_\nu]
\label{1.1f}
\end{equation}

\noindent is the curvature. We assume that the connection $a_\mu$ 
takes values in the algebra of U(N), 
with generators $T^A$ satisfying the trace normalization
\begin{equation}
tr(T^{A}T^{B})={\frac{1}{2}}\delta ^{AB}  \label{1.1a}
\end{equation}

\noindent and  the (anti)commutation relations
\begin{eqnarray}
\lbrack \,T^{A},T^{B}] &=&if^{ABC}T^{C}  \nonumber \\
\{T^{A},T^{B}\} &=&d^{ABC}T^{C}  \label{1.1b}
\end{eqnarray}

\noindent where $f^{ABC}$    and $d^{ABC}$  are assumed to be completely 
antisymmetric and completely symmetric respectively. 
  
The action (\ref{1.1}) is invariant under the infinitesimal gauge transformations
\begin{equation}
\label{1.2}
\bar\delta a_{\mu }\,=\, D_\mu \alpha\,\equiv \, \partial_\mu \alpha 
- i [ a_\mu , \alpha ]
\end{equation}

\noindent which closes in the algebra 

\begin{equation}
[\bar\delta_1,\bar\delta_2]\,a_\mu=\bar\delta_3\, a_\mu
\label{1.2a}
\end{equation}

\noindent with parameter composition rule given by

\begin{equation}
\alpha_3 = i[\alpha_2,\alpha_1]
\label{1.2b}
\end{equation}

The gauge structure displayed above leads to the definition  of the BRST differential 
$\bar s$ such that

\begin{eqnarray}
\bar sc&=&ic^2\nonumber\\
\bar s a_{\mu }&=&D_\mu c\nonumber\\
&\equiv&\partial_\mu c\ - i[a_{\mu} , c]\nonumber\\
\bar s\bar c &=& \gamma\nonumber\\
\bar s\gamma &=& 0
\label{1.6}
\end{eqnarray}

\noindent As $\bar s$ is an odd derivative acting from the right, 
it is easy to verify from the above definitions that it indeed is
 nilpotent. Naturally $c$ and $\bar c$ are grassmannian quantities with 
ghost numbers respectively $+1$ and $-1$.
$\bar c$ and $\gamma$ form a trivial pair necessary to implement the gauge fixing.
\medskip

The functional BRST quantization starts by defining the total action 

\begin{equation}
 S^{(0)} = S_0^{(0)} + S_1^{(0)}
 \label{1.3}
\end{equation}

\noindent where $S_0^{(0)}$ is given by (\ref{1.1}) and

\begin{equation}
 S_1^{(0)}=-2\,\,tr \,\bar s\,\int d^{4}x\, \bar c \left( -\frac{\gamma}{\beta}
+\partial_\mu a^\mu\right)
\label{1.3a}
\end{equation}

\noindent appropriated to fix the (Gaussian) Lorentz condition, is BRST exact. 
This assures that $S_1^{(0)}$ is  BRST invariant, due to the nilpotency of $\bar s$.
Since $\bar s f_{\mu\nu}=i[c,f_{\mu\nu}]$ according to (\ref{1.1f}) and 
(\ref{1.6}), it follows that $S_0^{(0)}$ is also BRST invariant. In (\ref{1.3a}), 
$\beta$ is a free parameter, as usual.

In general $ S_1^{(0)}$ can be written as
\begin{equation}
 S_1^{(0)}=S_{gh}^{(0)}+S_{gf}^{(0)}
\label{1.3b}
\end{equation}

\noindent with the ghost action given by

\begin{equation}
S_{gh}^{(0)}=-\,2\,tr\int d^{4}x\,\bar c M c
 \label{1.4}
\end{equation}

\noindent and the gauge fixing one by

\begin{equation}
S_{gf}^{(0)}=\,-2\,tr\int d^{4}x\,\gamma\left(-\frac{\gamma}{\beta} +{\cal F}\right)
 \label{1.5}
\end{equation}

\noindent where ${\cal F} = {\cal F} (a) $ is a gauge fixing function and 
$M={{\delta {\cal  F}}/{\delta\alpha }}$.
The Lorentz gauge condition corresponds to ${\cal F}_1=\partial_\mu a^\mu$, 
$M_1=\partial_\mu D^\mu$. The functional quantization is constructed by 
functionally integrating the exponential of $i\,S$ over all the fields, 
with appropriate measure and the external source terms in order to generate 
the Green's functions.

\medskip

The noncommutative version of this theory comes from replacing 
$y=\{a_\mu,\, c,\, \bar c,\,\gamma\}$  by the
noncommutative fields $Y=\{A_\mu,\, C,\,\bar C,\,\Gamma\}$ 
as well as the ordinary products of fields by  $\star$-Moyal products, defined through

\begin{equation}
\star\equiv exp\left(\frac{i}{2}\theta^{\mu\nu}\buildrel\leftarrow\over\partial_\mu 
\buildrel\rightarrow\over\partial_\nu \right)
 \label{1.5z}
\end{equation}

\noindent where $\theta^{\mu\nu}$ is a constant and antisymmetric matrix. 
The Moyal  product is associative and cyclic under the integral sign when 
appropriate boundary conditions are adopted. We also assume that the group 
structure is deformed by this product. In this way, the group elements are 
constructed from the exponentiation of elements of the algebra of U(N) by 
formally using $\star$ in the series, that is, 
$g=1+\lambda^A T^A+\frac{1}{2}\lambda^A T^A\star\lambda^B T^B+...$. 
General consequences of this deformation in field theories can be seen
in \cite{SW,REVIEW}. In particular, the appearance of $\theta^{0\,i} \ne 0\,$ 
breaks the unitarity of the corresponding quantum theory. Furthermore in a Hamiltonian
formalism this would imply higher order time derivatives which demand a non standard
canonical treatment. This last aspect does not show up here since we are using a 
Lagrangian formalism.

The noncommutative action corresponding to (\ref{1.1})  can be written as 

\begin{eqnarray}
S_0 &=& tr\int d^{4}x\,\Big(-\frac{1}{2}F_{\mu \nu }\star F^{\mu \nu } \Big)
\nonumber\\ \label{1.7}
\end{eqnarray}

\noindent where now

\begin{equation}
F_{\mu \nu } =\partial _{\mu }A_{\nu }-\partial _{\nu }A_{\mu}-i\,[A_{\mu } 
\buildrel\star\over, A_{\nu }] 
\label{1.8}
\end{equation}

As expected, the noncommutative gauge transformations 
$\delta A_{\mu }
=\partial_\mu \epsilon\ - i[A_{\mu} \buildrel\star\over, \epsilon]$
close in an algebra like (\ref{SW2a})
with composition rule for the parameters  given by

\begin{equation}
\epsilon_3=i[\epsilon_2\buildrel\star\over,\epsilon_1]
\label{1.3bb}
\end{equation}

The total action is given by 

\begin{equation}
 S=S_0 + S_1
 \label{1.3d}
\end{equation}

\noindent where 

\noindent

\begin{equation}
 S_1=-2\,\,tr  s\,\int d^{4}x\, \bar C 
\left( -\frac{\Gamma}{\beta}+\partial_\mu A^\mu\right)
\label{1.3e}
\end{equation}

\noindent The BRST differential $s$  is defined through

\begin{eqnarray}
s C&=&-iC\star C\nonumber\\
s A_{\mu }&=&D_\mu C\nonumber\\
&\equiv&\partial_\mu C\ - i[A_{\mu} \buildrel\star\over, C]\nonumber\\\ 
s\bar C&=&\Gamma\nonumber\\
s\Gamma&=&0
\label{1.9}
\end{eqnarray}

\noindent Obviously
both $S_0$ and $S_1$ are  BRST invariant.

The BRST Seiberg-Witten map is obtained from the condition (\ref{0.0}).
In this work we will consider only the expansion of noncommutative fields   
in terms of ordinary fields to first order in the noncommutative 
parameter $\theta$: 
$$Y[y] = y + Y^{(1)}[y] + O (\theta^2 )\,,$$
\noindent where $Y$ represents the noncommutative fields $A_\mu , C ,\bar C , \Gamma\,.$ 
Then we find from (\ref{0.0}) and (\ref{1.9})  that

\begin{eqnarray}
\bar s C^{(1)}+i\{c,C^{(1)}\}&=&\frac{1}{4}
\theta^{\alpha\beta}[\partial_\alpha c,\partial_\beta c]\nonumber\\
\bar s A^{(1)}_\mu+i[A^{(1)}_\mu, c ]&=&\partial_\mu C^{(1)}+
i[C^{(1)},a_\mu]\nonumber\\
&-&\frac{1}{2}\theta^{\alpha\beta}\{\partial_\alpha\,c,\partial_\beta a_\mu\}
\nonumber\\
\bar s\bar C^{(1)}&=&\Gamma^{(1)}\nonumber\\
\bar s \Gamma^{(1)}&=&0
\label{1.10}
\end{eqnarray}

\noindent The corresponding solutions for the ghost and the gauge field are

\begin{equation}
C^{(1)}\,=\,\frac{1}{4}\theta^{\mu \nu }\left\{\partial_\mu\,c,
a_\nu\right\} + 
\lambda_1\theta^{\mu \nu }\left[\partial_\mu\,c,
a_\nu\right]
 \label{1.11}
\end{equation}

\begin{eqnarray}
A^{(1)}_\mu &=& -\frac{1}{4}\theta^{\alpha\beta}
\left\{a_\alpha,\partial_\beta a_\mu
+f_{\beta\mu}\right\}\,+ 
\sigma\theta^{\alpha\beta}D_\mu f_{\alpha\beta}\nonumber \\
&+& {\lambda_1\over 2} \theta^{\alpha\beta}
{ D}_\mu [ a_\alpha , a_\beta ]
\,,
\label{1.12}
\end{eqnarray}

\noindent where $\sigma$ and $\lambda_1$ are arbitrary constants.

It is important to remark that when we extend the space of fields
in order to implement BRST quantization we could in principle find solutions
for $ C^{(1)}$ and $A^{(1)}_\mu $ depending on $\bar c $ and $\gamma$. 
However there are two additional conditions to be satisfied, besides 
eqs. (\ref{1.10}): the ghost number and the dimension of all first order 
corrections must be equal to those of the corresponding zero order field. 
It can be checked that there are no other possible contributions to the 
solutions (\ref{1.11}) and (\ref{1.12}) involving $\bar c $ and $\gamma$ 
and satisfying these criteria.

For the trivial pair we find

\begin{eqnarray}
{\bar C}^{(1)} &=& \theta^{\alpha\beta} \, H_{\alpha\beta}  
\nonumber\\
\Gamma^{(1)}&=& \theta^{\alpha\beta} \,\bar s \,H_{\alpha\beta}  
\label{1.13}
\end{eqnarray}

\noindent where $H_{\alpha\beta}$ is a function of the fields $a_\mu , c, \bar c $ 
and $\gamma $ with ghost number $= -1\,$ (taking the convention that the ghost numbers 
of  $c$ and $\bar c$ are $ +1$ and $- 1$ respectively).
Note that the sum of the mass dimensions of $ c$ and $  \bar c $ is 2. 
Then the mass dimension of $H_{\alpha\beta}$ will be the
dimension of $ \bar c$ plus 2.  Considering these points we find that the general form
for this quantity is 

\begin{equation} 
\label{nova}
 H_{\alpha\beta} \,=\, \omega_1 \,\bar c \, a_{_{[\alpha}} a_{_{\beta]}} +
\omega_2 \,\bar c \, \partial_{_{[\alpha}} a_{_{\beta]}} +
\omega_3\, \partial_{_{[\alpha}} \bar c \, a_{_{\beta]}} 
\end{equation}

\noindent where $\omega_1 \,,\,\omega_2\,$ and $\omega_3$ are arbitrary parameters. 
From eqs. (\ref{1.6}), (\ref{1.13}) and (\ref{nova}) we see that 

\begin{eqnarray}
\Gamma^{(1)}&=& \theta^{\alpha\beta} \Big( 
\omega_1 \,\gamma \, a_{_{\alpha}} a_{_{\beta}} +
\omega_2 \,\gamma \, \partial_{_{\alpha}} a_{_{\beta}} +
\omega_3\, \partial_{_{\alpha}} \gamma \, a_{_{\beta}} 
\nonumber\\
&+& \omega_1 \,\bar c \,(  D_{_{\alpha}}\,c \,a_{_{\beta}} +
\, a_{_{\alpha}} D_{_{\beta}}\,c\,)
+ \omega_2 \, \bar c  \, \partial_{_{\alpha}} D_{_{\beta}}\,c\nonumber\\
&+&
\omega_3\, \partial_{_{\alpha}} \bar c \, D_{_{\beta}} \,c\,\Big)\,\,.
\label{nova2}
\end{eqnarray}

The usual Seiberg-Witten map is defined for the $S_0$ (gauge invariant) 
part of the action. When we gauge fix by including $S_{gh}$ and $S_{gf}$
we find a total action that is no more gauge invariant but rather 
BRST invariant. This poses a non trivial problem of whether it would still
be possible to relate noncommutative and ordinary gauge fixed theories
by a SW map.

Let us consider the Lorentz gauge condition appearing in (\ref{1.3e}):

\begin{equation}
{\cal F}_1 = \partial^\mu A_\mu \,=\, 0
\label{1.13a}
\end{equation} 

\noindent 
If we use the Seiberg-Witten map found above we see that this condition would 
correspond to 

\begin{eqnarray}
\partial^\mu a_\mu &=& 0 \label{1.13b}\\
\partial^\mu A^{(1)}_\mu [a_\mu] &=& 0 \,\,.
\label{1.13c}
\end{eqnarray}

\noindent That means, besides the ordinary Lorentz condition (\ref{1.13b}), 
we find the additional  non linear conditions on $a_\mu $ from (\ref{1.12})
and (\ref{1.13c}). 
If we were adopting in our expansions terms up to order $N$ in $\theta$, we would 
find a set of conditions 
$\partial^\mu A^{(n)}_\mu [a_\mu]=0$, $n=0,1,..,N$. 
So it seems that the condition 
$  \partial^\mu A_\mu \,=\, 0\,$ is not compatible with the solution (\ref{1.12}), 
and its higher order extensions,
for the SW mapping. 

Alternatively we could choose in the noncommutative theory the non trivial 
non linear gauge condition (to first order in $\theta$ )

\begin{eqnarray}
{\cal F}_2 &=& \partial^\mu A_\mu +\frac{1}{4}\theta^{\alpha\beta} \partial^\mu 
\left\{A_\alpha,\partial_\beta A_\mu
+F_{\beta\mu}\right\}\nonumber \\ 
&-& \sigma\theta^{\alpha\beta} \partial^\mu 
D_\mu F_{\alpha\beta}
\,-\, {\lambda_1\over 2} \theta^{\alpha\beta}
 \partial^\mu { D}_\mu [ A_\alpha , A_\beta ]
\nonumber\\
&=& \,\,0\,\,.
\label{1.14}
\end{eqnarray}

\noindent If we assume the map (\ref{1.12}) to hold, this gauge fixing 
condition would correspond just to the ordinary Lorentz condition
(\ref{1.13b}) to first order in the noncommutative parameter $\theta$.
We see that if we impose the Lorentz gauge in one of the theories, the SW 
map would lead to a somehow complicated and potentially inconsistent gauge in the other.

Instead of just considering the map of the gauge fixing
functions, a more general approach is to consider the behaviour of the action $S_1$  
eq. (\ref{1.3e}) under the SW map. This action  can be written, to first order in the 
noncommutative parameter, as

\begin{equation}
 S_1=S_1^{(0)}+S_1^{(1)}\,\,,
\label{1.15}
\end{equation}

\noindent where $S_1^{(0)}$ is given by (\ref{1.3a}) and

\begin{eqnarray}
 S_1^{(1)}&&=-2\,\,tr\,\bar s\,\int d^{4}x\,
\Big( \bar C^{(1)}( -\frac{\gamma}{\beta}+\partial_\mu a^\mu)
\nonumber\\
&&+\,\,\bar c\,\,(-\frac{\Gamma^{(1)}}{\beta}+\partial_\mu A^{(1)\mu})\Big)\,\,.
\label{1.16}
\end{eqnarray}

Note that the condition ${\cal F}_1 = \partial^\mu A_\mu \,=\, 0$ in the noncommutative  
theory would be effectively mapped into  ${\cal F}_3  = \partial^\mu a_\mu \,=\, 0$ 
if $S_1^{(1)}$ could vanish. 
In order to see if this is possible we introduce (\ref{1.12}) and (\ref{nova2}) 
in (\ref{1.16}) and examine the terms with the same field content.
The part of  $S_1^{(1)}$ independent of the ghost sector is

\begin{eqnarray}
S_{1\,no\,ghost}^{(1)} &=&-2\,\,\theta^{\alpha\beta}\, tr\,\int d^{4}x\,
\Big[ \Big( \partial_\mu a^\mu -  {2 \gamma\over \beta} \Big) 
\nonumber\\
&\times& \Big(
\omega_1 \,\gamma \, a_{_{\alpha}} a_{_{\beta}}
+ \omega_2 \,\gamma \, \partial_{_{\alpha}} a_{_{\beta}} +
\omega_3\, \partial_{_{\alpha}} \gamma \, a_{_{\beta}} \Big)
\nonumber\\
&+&\gamma \partial^\mu  \Big(
-\frac{1}{4} \left\{a_{_\alpha},\partial_{_\beta} a_{_\mu}
+f_{_{\beta\mu}}\right\}\,+ 
\sigma D_\mu f_{_{\alpha\beta}}\nonumber \\
&+& {\lambda_1\over 2} { D}_\mu [ a_{_\alpha} , a_{_\beta} ]\Big)
\Big]\,.
\label{nova3}
\end{eqnarray}

The terms linear in the connection $a_\mu$ in the integrand 
are 

\begin{equation}
- {2 \gamma\over \beta} \Big( 
\, \omega_2 \,\gamma \, \partial_{_{[\alpha}} a_{_{\beta]}} +
\omega_3\, \partial_{_{[\alpha}} \gamma \, a_{_{\beta]}} \Big)
+ 2 \sigma \gamma \Box 
\partial_{_{[\alpha}} a_{_{\beta\,]}} \,
\end{equation}

\noindent which will vanish modulo total differentials only if $\sigma = 0$ and
$\omega_3 = 2 \omega_2 $. Using these results the quadratic part in $a_\mu $
of the integrand can be written as 

\begin{eqnarray}
- {2\over \beta}\, \gamma^2  \omega_1 \, a_{_{[\alpha}} a_{_{\beta\,]}} 
\,-\, \gamma \Big( 
 \omega_2 \,(  \,\, \partial_{_{[\alpha}} \, a_{_{\beta\,]}} \, \partial_\mu a^\mu
+2 \,\, a_{_{[\beta}} \partial_{_{\alpha ]}} \partial_\mu a^\mu \,)
\nonumber\\
+  {1\over 4}
\partial^\mu (  2 \{ a_{_{[\alpha}} ,  \partial_{_{\beta\,]}} \, a_\mu \}
+  
\{ \partial_\mu  a_{_{[\alpha}} ,  a_{_{\beta ]}} \}
\,- \,4 \lambda_1 \, \partial_\mu ( a_{_{[\alpha}} \, a_{_{\beta ]}}) )\Big)\,
\nonumber\\ 
\end{eqnarray}  

\noindent modulo a total differential. The quadratic term 
in $\gamma\,$ can be set to zero choosing $\omega_1 = 0$.
However, there is no choice for $\lambda_1$ and 
$\omega_2$ which makes the linear terms in $\gamma$ in the expression above vanish
or be written as a total derivative. 
Since the terms from the ghost sector can not cancel the above ones, 
we conclude that $S_1^{(1)}$ can not vanish.
So it is not possible to relate the Lorentz gauge conditions in the ordinary 
and noncommutative theories by the SW map.
Anyway, quantization can be consistently implemented in both theories
as pointed out  in the end of section~{\bf I}.
We will see in section {\bf IV} that for the gauge invariant Proca model it is 
possible to find a particular BRST-SW map relating the gauge fixing unitary 
conditions for both sectors.

\section{The non-Abelian Proca model}
\renewcommand{\theequation}{3.\arabic{equation}}
\setcounter{equation}{0}

Before considering the gauge invariant noncommutative $U(N)$ Proca model, 
it will be useful to briefly present some basic properties of the ordinary 
Abelian Proca theory with action  

\begin{equation}
S_0^{(0)} =\,\int d^{4}x\,\left(\,-\frac{1}{4}f_{\mu \nu } f^{\mu \nu } +\frac{1}{2}
m^2 a_\mu a^\mu \right)\,\,.
 \label{2.1}
\end{equation}

\noindent Here $a_\mu$ represents the massive Abelian vector field.
Variation of (\ref{2.1}) with respect to $a_\mu$ gives the equation of motion

\begin{equation}
\partial_\mu f^{\mu \nu} +m^2  a^\nu=0
 \label{2.2}
\end{equation}

\noindent which, by symmetry, implies that

\begin{equation}
\partial_\mu a^\mu=0
 \label{2.3}
\end{equation}

\noindent Substituting (\ref{2.3}) into (\ref{2.2}) we find  
that the vector field satisfies a massive Klein-Gordon equation, as expected:

\begin{equation}
(\Box+m^2) a_\mu=0
 \label{2.4}
\end{equation}

This model can be described in a gauge invariant way by introducing
a compensating field $\lambda\,$. This kind of mechanism is useful, for instance,
when calculating anomalies\cite{CF,ABH}.  In this case, the action (\ref{2.1}) 
is replaced by

\begin{equation}
S_0^{(0)} = \int d^{4}x \left(\,-\frac{1}{4}f_{\mu \nu } f^{\mu \nu } +\frac{1}{2}
m^2 (a_\mu-\partial_\mu\lambda)( a^\mu- \partial^\mu\lambda)\right) .
\label{2.5}
\end{equation}

\noindent This action is invariant under the local transformations 
$\delta a_\mu= \partial_\mu\alpha$ and $\delta\lambda=\alpha$. 
The equations of motion for $a_\mu$ and $\lambda$ are 

\begin{eqnarray}
\partial_\mu f^{\mu \nu} +m^2 ( a^\nu-\partial^\nu\lambda)&=&0\nonumber\\
\partial_\mu( a^\mu-\partial^\mu\lambda)&=&0
\label{2.6}
\end{eqnarray}

Note that applying $\partial_\nu$ to the first equation, one reobtains 
the second one. In this case (\ref{2.3}) does not come from 
the equations of motion.
However as now the model is gauge invariant, we must impose a 
gauge fixing function. We may choose for instance one of the gauge fixing functions
${\cal F}_1=\partial_\mu a^\mu$, ${\cal F}_2=\Box \lambda$ 
or ${\cal F}_3=\lambda$ in order to recover the original Proca theory. 

We now consider the non-Abelian generalization of this model. 
Now $a_\mu$ takes values in the algebra of U(N), 
exactly as in the Yang-Mills case of the previous section.

In place of (\ref{2.2}) one finds 

\begin{equation}
D_\mu f^{\mu \nu} +m^2  a^\nu=0
\label{2.10}
\end{equation}

Applying $D_\nu $ to this equation and using the property

\begin{equation}
[D_\mu,D_\nu]y=i[f_{\mu\nu}\,,\,y]      
\label{2.11}
\end{equation}

\noindent where $y$ is any arbitrary function with values in the algebra, 
we find as in the Abelian case that

\begin{equation}
D_\mu a^\mu \,=\, \partial_\mu a^\mu \,=\,0\,\,.
\label{2.12}
\end{equation}

However, the equations of motion present non linear terms.
Using the "Lorentz  identity"  (\ref{2.12}) in eq. (\ref{2.10})
we obtain the equations of motion for $a^\mu $

\begin{equation}
(\Box+m^2) a^\rho-i[a_\mu,\partial^\mu a^\rho+f^{\mu\rho}]=0
 \label{2.13}
\end{equation}

\noindent which is no longer a Klein Gordon equation.

It is worth to mention that contrarily to the Abelian case, the non-Abelian 
Proca model is not renormalizable, although the divergencies at one-loop 
level cancel  in an unexpected  way \cite{Itzykson},\cite{Zinn-Justin}.  
Nonrenormalizability is, in any way,
an almost general property of noncommutative field theories\cite{REVIEW}.

The next step would be to consider the gauged version of this non-Abelian model. 
This will be done in the noncommutative context in the next section.
We will also discuss there the gauge fixing procedure and the BRST formalism. 
  
\section{The gauge invariant noncommutative $U(N)$ Proca model}
\renewcommand{\theequation}{4.\arabic{equation}}
\setcounter{equation}{0}

The gauge invariant version of the U(N) noncommutative Proca model can be written 
as\cite{ABF}

\begin{eqnarray}
S_0 &=& tr\int d^{4}x\,\Big( -\frac{1}{2}F_{\mu \nu } \star F^{\mu \nu } 
\nonumber\\
&+& m^2 (A_\mu-B_\mu)\star ( A^\mu - B^\mu)\Big)
 \label{3.1}
\end{eqnarray}

\noindent where  the curvature is again given by eq. (\ref{1.8}) and

\begin{equation}
B_\mu\equiv -i\,\partial_\mu G \star G^{-1}
\label{3.2}
\end{equation}

In the above expressions $G$ is an element of the noncommutative U(N)  group
and $B_\mu$ is a "pure gauge" vector field in the sense that its curvature, 
analogous to (\ref{1.8}), vanishes identically.
Note that $B_\mu$ is the noncommutative $ U(N)$ version of  
$\partial_\mu \lambda$ discussed in the previous section.
We assume that the algebra generators satisfy, as in the Yang-Mills case, the trace 
normalization and (anti)commutation relations (\ref{1.1a}) and (\ref{1.1b}).

By varying action (\ref{3.1}) with respect to $A_\mu$ and  $G$, we get the 
equations of motion

\begin{eqnarray}
D_\mu F^{\mu \nu} +m^2 ( A^\nu-B^\nu)&=&0\label{3.3a} \\
\bar D_\mu(  A^\mu-B^\mu)&=&0
\label{3.3b}
\end{eqnarray}

\noindent where we have defined the two covariant derivatives

\begin{eqnarray}
D_\mu X&=&\partial_\mu X-i[A_\mu\buildrel\star\over, X]\nonumber\\
\bar D_\mu X&=&\partial_\mu X-i[B_\mu\buildrel\star\over, X]
\label{3.4}
\end{eqnarray}

\noindent for any quantity $X$ with values in the algebra. By taking the covariant 
divergence  of equation (\ref{3.3a}) and using the noncommutative Bianchi 
identity

\begin{equation}
[D_\mu,D_\nu]X=i[F_{\mu\nu}\buildrel\star\over,X]      
\label{3.5}
\end{equation}

\noindent one finds

\begin{equation}
D_\mu(A^\mu-B^\mu)=0    
\label{3.6}
\end{equation}

\noindent which is equivalent to equation (\ref{3.3b}), as can be verified. 
Actually, we can rewrite (\ref{3.3b}) or (\ref{3.6}) in the convenient form

\begin{equation}
\partial_\mu A^\mu=D_\mu B^\mu\,\,.
\label{3.7}
\end{equation}

\noindent So we see that if we choose, among the possible gauging fixing 
functions, ${\cal F}_1=\partial_\mu A^\mu$ or ${\cal F}_4=G-1$, the compensating field is 
effectively eliminated and the Lorentz condition is implemented.
\medskip

Action (\ref{3.1}) is invariant under the
infinitesimal gauge transformations 

\begin{eqnarray}
\delta A_{\mu }&=&D_\mu \epsilon\nonumber\\
\delta G&=&i\epsilon \star G
\label{3.8}
\end{eqnarray}

\noindent which also implies that 

\begin{eqnarray}
\delta B_{\mu }&=&\bar D_\mu \epsilon\nonumber\\
\delta F_{\mu \nu} &=& i[ \epsilon\buildrel\star\over,F_{\mu \nu}] 
\label{3.9}
\end{eqnarray}

\noindent 
As expected, the noncommutative gauge transformations listed above 
close in an algebra as (\ref{SW2a}), for the fields
$A_\mu$,  $G$ , $B_\mu$ or $F_{\mu\nu}$.
The composition rule for the parameters is as (\ref{1.3bb}).
 
The associated BRST algebra obtained again by introducing the trivial pair 
$\bar C$ and $\Gamma$,  corresponds to the Yang-Mills algebra of
eq. (\ref{1.9}) plus the transformation of the compensating field

\begin{equation}
s G \,=\,iC \star G\,\,,
\label{3.12}
\end{equation}

\noindent that corresponds to

\begin{equation}
s B_{\mu }\,=\, \bar D_\mu C = \partial_\mu C - i [ B_\mu \buildrel\star\over, C]\,\,.
\label{3.13}
\end{equation}

A gauge fixed action is constructed from (\ref{3.1}) 
by adding $S_{gh} $ and $S_{gf}$ from eqs. (\ref{1.4}) and (\ref{1.5})
as in the Yang-Mills case.

For the Lorentz gauge we choose as in the Yang-Mills case 
${\cal F}_1=\partial_\mu A^\mu$, 
$M_1=\partial_\mu D^\mu$ and we get formally the same ghost and gauge fixing 
actions.
The complete action is BRST invariant, as expected.

For the unitary gauge corresponding to the choice ${\cal F}_4=G-1$,
we find the ghost and gauge fixing actions

\begin{eqnarray}
S_1&=&S_{gh}+S_{gf}\nonumber\\
&=& -2 tr\,s\,\int d^{4}x\,\bar C \,(G-1)
 \label{3.14}
\end{eqnarray}
\noindent which is obviously BRST invariant. Since  $s\,S_0=0$, as can be 
verified, it follows the invariance of the complete action.

Let us now build up the BRST Seiberg-Witten map for this model.
We start again by imposing \cite{AF2} that $ s Y=\bar s Y[y] $ submitted to the condition 
$ Y[y]_{\mid_{\theta=0}}=y $ that lead us again to equations (\ref{1.10})
and also to

\begin{eqnarray}
\bar s G^{(1)}-icG^{(1)}&=&- \frac{1}{2}\theta^{\mu\nu}\partial_\mu\,
c\partial_\nu\,g + i  C^{(1)} \,g\nonumber\\
\bar\delta B^{(1)}_\mu+i[B^{(1)}_\mu , c]&=& \partial_\mu C^{(1)}+
i[C^{(1)},b_\mu]\nonumber\\
&-&\frac{1}{2}\theta^{\alpha\beta}\{\partial_\alpha\,c,\partial_\beta b_\mu\}
\label{4.1}
\end{eqnarray}

The solutions of these equations are again  obtained by searching for all the
possible contributions  with the appropriate dimensions 
and Grassmaniann  characters. However now we have the extra compensating field
$b_\mu$.
So the solutions for $ A^{(1)}_\mu$ and $C^{(1)}$ are no longer the Yang-Mills 
solutions (\ref{1.11}) and (\ref{1.12}) but rather

\begin{eqnarray}
C^{(1)}&=&\frac{1}{4}(1 - \rho )\theta^{\mu \nu }\left\{\partial_\mu\,c,
a_\nu\right\} + 
\lambda_1\theta^{\mu \nu }\left[\partial_\mu\,c,
a_\nu\right]\nonumber\\
& +& \frac{1}{4}\rho \theta^{\mu \nu }\left\{\partial_\mu\,c,
b_\nu\right\}
+ \lambda_2\theta^{\mu \nu }\left[\partial_\mu\,c,
b_\nu\right]
\label{4.2}
\end{eqnarray}

\begin{eqnarray}        
A^{(1)}_\mu &=& - {1-\rho\over 4} \theta^{\alpha\beta}
\left\{a_\alpha,  \partial_\beta a_\mu + f_{\beta\mu} \right\}
+ \sigma \theta^{\alpha\beta} D_\mu f_{\alpha \beta}\nonumber\\
&+& {\lambda_1\over 2} \theta^{\alpha\beta} D_\mu [ a_\alpha , a_\beta ]
+ \frac{\rho}{4}\theta^{\alpha\beta}\left\{b_\alpha,  D_\mu b_\beta - 
2 \partial_\beta a_\mu\right\} \nonumber\\
&+& \lambda_2 \theta^{\alpha\beta} D_\mu \Big( b_\alpha b_\beta \Big)
\label{4.3}
\end{eqnarray}

\noindent  In these equations $\lambda_1 $,  $\lambda_2$ , $\rho$ and $\sigma$ 
are arbitrary constants. 
These solutions are a generalization of those in ref. \cite{ABF}.

For the compensating field we find

\begin{eqnarray} 
G^{(1)}&=&-\frac{1}{2}(1-\rho)\theta^{\alpha\beta}
a_\alpha\left(\partial_\beta g-\frac{i}{2}a_\beta g\right)\nonumber\\
&+&i\lambda_1\theta^{\alpha\beta}a_\alpha a_\beta g
+\gamma \theta^{\alpha\beta} f_{\alpha\beta}g \nonumber\\
&+&
i(\lambda_2-\frac{\rho}{4})\theta^{\alpha\beta}b_\alpha b_\beta g
+
O(\theta^2)
\label{SW14}
\end{eqnarray}

\noindent with arbitrary $\gamma$. 

Now we may consider the compatibility of the BRST-SW map and the unitary gauge fixing, 
in the same spirit of what was discussed in the last section. 
First we observe that the condition  ${\cal F}_4=G-1$ is mapped in $g-1$ plus 
complicated first order  corrections in $\theta$.
However, if we choose the parameters of the map as

\begin{eqnarray}
\lambda_1 &=& {1\over 4}( \rho - 1) \nonumber\\
\gamma &=& 0
\end{eqnarray}

\noindent we find that the conditions $ g = 1 $ and $ G =1 $ are equivalent
to first order in $\theta$, as can be verified from (\ref{SW14}). So that for 
this particular solution of the map, the unitary gauge in the noncommutative 
theory is mapped in the ordinary unitary gauge in the mapped theory.

Additionally, if we choose $\rho = 1$ implying $\lambda_1 = 0$ and also $\sigma =0$
we find that in the unitary gauge $ A^{(1)}_\mu = 0$. 
So that $ A_\mu $ is mapped just in $a_\mu$.  
In this  case we find that the non gauge invariant noncommutative 
$U(N)$ Proca theory is recovered in the unitary gauge and at 
the considered order in $\theta$.
 
\section{Conclusion}
We have investigated the extension of the Seiberg-Witten map to ghosts, antighosts
and auxiliary fields in the BRST gauge fixing procedure. 
Two non-Abelian gauge models that present different 
behaviors under the BRST-SW map of the gauge fixing conditions have been considered. 
For the Yang-Mills theory we found that it is not possible to map the Lorentz
gauge condition in the noncommutative and ordinary theories.
On the other hand, for the gauged $U(N)$ Proca model we were able to find a 
SW map that relates the unitary gauge fixing in the noncommutative and ordinary 
theories.
It would be interesting to investigate how the BRST-SW map act in other 
gauge fixing conditions and models. 

\medskip

Acknowledgment: This work is supported in part by CAPES and CNPq (Brazilian
research agencies).


\begin{thebibliography}{a}
\bibitem[a]{a} e-mail: amorim@if.ufrj.br

\bibitem[b]{b} e-mail: boschi@if.ufrj.br

\bibitem[c]{c} e-mail: braga@if.ufrj.br 


\bibitem{SW}  N. Seiberg and E. Witten, JHEP 09 (1999) 32.

\bibitem{AK} T.~Asakawa and I.~Kishimoto,
JHEP 11 (1999) 024.

\bibitem{Wess} B. Jurco, J. Moller, S. Schraml, P. Schupp and J. Wess, Eur.
Phys. J. C 21 (2001)383; D. Brace, B. L. Cerchiai, A. F. Pasqua, U. Varadarajan
and B. Zumino, JHEP 0106 (2001) 047.

\bibitem{REVIEW} For a review see  R.J. Szabo, Phys. Rep. 378 ( 2003) 207 and 
references therein.


\bibitem{ABF} R. Amorim, N. R. F. Braga and C. N. Ferreira, Phys. Lett. B591 (2004) 181.

\bibitem{AFG} R. Amorim, C. N.  Ferreira and C. Godinho, 
Phys. Lett. B593 (2004) 203.

\bibitem{AF2} R. Amorim and F. A. Farias, Phys. Rev. D69 (2004) 045013. See also
O. F. Dayi, Phys. Lett. B481 (2000) 408.

\bibitem{Po} J.Polchinski, Phys. Rev. Lett. 75 (1995) 4724.

\bibitem{HV} C. Hofman , E. Verlinde, J. High Energy Phys. 12 (1998) 10.

\bibitem{CH1} C.-S.Chu , P.-M. Ho, Nucl. Phys. B550 (1999) 151.

\bibitem{Heisenberg} W. Pauli, Scientific Correspondence, 
Vol II p.15, Ed. Karl von Meyenn, Springer-Verlag, 1985.

\bibitem{Snyder} H.S. Snyder \textit{Quantized Space-Time}, 
Phys. Rev. 71 (1) (1947) 38.

\bibitem{Bonora} L. Bonora and L. Salizzoni, Phys. Lett. B 504 (2001) 80; A.
Armoni, Nucl Phys. B593 (2001) 229; M.M. Sheikh-Jabbari, 
Nucl.Phys.Proc.Suppl.108 (2002) 113-117 .

\bibitem{AF} R. Amorim and F. A. Farias, Phys. Rev. D 65 (2002) 065009.

\bibitem{CF} B de Wit and M. T. Grisaru, ``Compensating Fields and Anomalies" in
Quantum Field Theory and Quantum Statistics, Vol.2,   eds. I.A.Batalin,
C.J.Isham and  G.A. Vilkovisky, Adam Hilger, 1987.

\bibitem{ABH} R. Amorim, N. R. F. Braga and M. Henneaux, 
Phys. Lett. B436 (1998) 125.

\bibitem{FV} M. Henneaux and C. Teitelboim,``Quantization of Gauge Systems", 
Princeton University Press, 1992. 

\bibitem{Itzykson} C. Itzykson  and J-B Zuber,  
``Quantum Field Theory", McGraw-Hill Book Company, 1980.

\bibitem{Zinn-Justin} J. Zinn-Justin,  ``Quantum Field Theory 
and Critical Phenomena", Oxford Science Publications, 1996.

\end{thebibliography}
\end{document}